\documentclass[article, aps,preprint,preprintnumbers,amsmath,amssymb,nofootinbib]{revtex4-1}
\usepackage{epsfig}
\usepackage{dcolumn}
\usepackage{graphicx}
\graphicspath{ {images/} }
\usepackage{bm}
\usepackage{hyperref}
\usepackage{color}  
\usepackage{setspace}
\usepackage{appendix}
\usepackage{subcaption}
\usepackage{mathtools}

\newcommand{\Alfven}{Alfv{\'e}n\hspace{0.1cm}}
\newcommand{\Alfvenic}{Alfv{\'e}nic\hspace{0.1cm}}

\newcommand{\vel}{\mathbf{u}}
\newcommand{\B}{\mathbf{B}}

\newcommand{\pd}{\partial }

\definecolor{Gray}{gray}{0.85}
\definecolor{LightCyan}{rgb}{0.88,1,1}
\begin{document}

\title{Effect of In-Plane Shear Flow on the Magnetic Island Coalescence Instability}

\author{Jagannath Mahapatra\textsuperscript{1,2}}
\email{jaga.physics94@gmail.com; jagannath.mahapatra@ipr.res.in}
\author{Arkaprava Bokshi\textsuperscript{1}}
\author{Rajaraman Ganesh\textsuperscript{1,2}}
\email{ganesh@ipr.res.in} 
\author{Abhijit Sen\textsuperscript{1,2}}

\affiliation{\textsuperscript{1}Institute for Plasma Research, Gandhinagar, Gujarat-382 428, India}
\affiliation{\textsuperscript{2}Homi Bhabha National Institute, Mumbai, Maharashtra-400 094, India}
%

\begin{abstract}
Using a 2D Viscoresistive Reduced MagnetoHydroDynamic (VR-RMHD) model, the magnetic island coalescence problem is studied in the presence of in-plane, parallel shear flows. {Extending the analytical work of Waelbroeck \emph{et al.}, Phys. Plasmas {\bf{14}}, 022302 (2007) and Throumoulopoulos \emph{et al.}, J. Phys. A: Math. Theor. {\bf{42}}, 335501 (2009)} in the sub-\Alfvenic flow shear regime for Fadeev equilibrium, the super-\Alfvenic regime is studied for the first time numerically. A wide range of values of shear flow amplitudes and shear scale lengths have been considered to understand the effect of sub-\Alfvenic and super-\Alfvenic flows {{on the coalescence instability and its nonlinear fate}}. We find that for {{flow}} shear length scales greater than the magnetic island size, the maximum reconnection rate decreases monotonically from sub-\Alfvenic to super-\Alfvenic flow speeds. For scale lengths smaller than the island size, the reconnection rate decreases up to a critical value $v_{0c}$, beyond which, the shear flow is found to destabilize the islands. The value of $v_{0c}$ decreases with a decrease in the value of shear flow length scale. Interestingly, for our range of parameters, we find suppression of the Kelvin-Helmholtz instability in super-\Alfvenic flows even when the shear scale length is smaller than the island width. Observation of velocity streamlines shows that the plasma circulation inside the islands {{has a stabilizing influence}} in strong shear flow cases. Plasma circulation is also found to be responsible for the decrease in upstream velocity, causing less pile-up of magnetic flux on both sides of the reconnection sheet.
\end{abstract}
\maketitle

\section{Introduction}
\indent
Magnetic reconnection (MR) is a fundamental phenomenon in dynamically evolving plasma systems that is responsible for relaxing the magnetic field topology by converting the magnetic energy into kinetic energy. This phenomenon heats up the plasma \cite{Biskamp_2000_MRbook, priest_forbes_2000_MRbook} and is frequently observed in both laboratory and space plasma. {{Examples include saw-tooth crash driven by tearing mode instability (TMI) in tokamaks \cite{Wesson1987}, solar flares, coronal mass ejection, magnetospheric substorms, etc. (see \cite{Zweibel2009} and references therein)}}.  
{{The most favorable location for two-dimensional MR to occur is the region of thin current sheet that develops at the $X$-type magnetic null points ($X$-type null lines in 3D are topologically unstable) generated through MHD instabilities, e.g. the TMI \cite{Lapenta2008}, plasmoid instability \cite{Hsseinpour2018} and island coalescence instability \cite{Stainner2013, Stainner2017, rmhd1,rmhd2,rmhd3, Pritchett-Wo-1979, Pritchett1992, Karimabadi2011, finn_Kaw1977, Knoll_chacon2006, Bard2018,Makwana2018}, to name a few}}. Any change in plasma parameters around these local $X$-points is known to affect the {{entire}} process of MR. A large number of numerical simulations have been conducted to understand the role of various parameters on MR, such as the effect of non-uniform resistivity \cite{Baty2007}, presence of strong guide field \cite{Stainner2017}, presence of asymmetric magnetic field and density on the upstream side of reconnection current sheet \cite{Doss2015,Eastwood2013} and more. Additionally, as the plasma bulk flow is common in fusion plasma experiments (e.g., generated indirectly by neutral beam injection or wave heating/current driving mechanisms) as well as in space plasma environment (e.g., caused due to plasma jets from astronomical bodies, stellar wind, etc.), it can also affect MR in many important ways by altering the upstream and downstream flow pattern.\\
\indent As is well known, the magnetic field can also affect the flow dynamics. For example, the presence of magnetic field parallel to shear flow with velocity discontinuity generates Kelvin-Helmholtz instability (KHI) if the difference in velocity across the discontinuity layer is greater than twice the \Alfvenic velocity ($v_A$) \cite{Keppens1999, SChandrasekhar_KHI_book}. In the past, most of the research work on MR in the presence of shear flows have used two kinds of MHD equilibrium: the Harris current sheet equilibrium \cite{Harris1962} and the Fadeev or magnetic current island equilibrium \cite{Fadeev1965} (used interchangeably throughout this work). The Harris type equilibrium has a long thin current sheet that separates two regions of anti-parallel magnetic field. In the absence of shear flow, when the current sheet becomes unstable to TMI, it breaks down into large magnetic islands through MR. However, in the presence of in-plane parallel \cite{Chen-Morrison1990,Ofman-Morrison1993,Cassak2011,Hsseinpour2018} or anti-parallel \cite{Ofman-Morrison1993,Chen1997} shear flows, the rate of island formation, equivalently the reconnection rate, reduces monotonically with an increase of shear flow amplitude up to a critical value, after which the TM mode transits to the KH mode. This critical shear flow amplitude for a resistive MHD model is $v_A$. The inclusion of Hall physics reduces the critical flow amplitude to sub-\Alfvenic values \cite{Chacon2003} enabling the KHI and TMI to couple. However, as stronger shear flows suppress the growth and size of islands, it becomes difficult to study the TM generated magnetic islands in the presence of \Alfvenic and super-\Alfvenic flows. One of the drawbacks of using the Harris type equilibrium is that the initiation of TMI driven MR requires either an external driving force (e.g. Newton's challenge problem \cite{Birn_Hesse2007}) or the current sheet aspect ratio needs to be very large (thin and long current sheet). {{Moreover, in natural reconnecting systems, thin current sheets develop dynamically at the $X$-points. However, the pre-formed current sheet structure of Harris equilibrium is less suitable to address reconnection physics (see Ref. \cite{Stainner2017} for further details). Interestingly, Fadeev equilibrium \cite{Fadeev1965} has such features inherently built into it. As Fadeev equilibrium contains a 1D chain of current filaments separated by $X$-type magnetic null points, in the presence of finite dissipation, the force of attraction between the parallel current strands brings them closer leading to the formation of a thin reconnection current sheet at the $X$-point and drives the MR. This self-driven mechanism for MR is also one of the advantages of this equilibrium \cite{Stainner2017}}}. Mutually attracting magnetic islands finally coalesce to form a bigger island.\\
\indent In the past, several attempts have been made to understand the island coalescence problem. For example, stability analysis of a 2D island configuration has been studied analytically by Finn and Kaw \cite{finn_Kaw1977}. To understand the observed fast MR time scale, 2D island coalescence problem has been studied numerically using resistive MHD \cite{rmhd1, rmhd2, rmhd3, Pritchett-Wo-1979}, Hall-MHD \cite{Knoll_chacon2006,Bard2018}, kinetic \cite{Pritchett1992,Karimabadi2011}, ten moment two-fluid \cite{Stainner2017} and hybrid \cite{Makwana2018} simulation models. In the above-mentioned body of work, the role of several external parameters such as system size, guide field strength on the properties of MR has been addressed. However, the effect of shear flow on island coalescence has not been studied so far. Furthermore, Fadeev equilibrium has well developed current islands to couple with both sub-\Alfvenic and super-\Alfvenic shear flows. Previously, several analytical studies have reported the shear flow effects on the Kelvin-Stuart \cite{finn_Kaw1977} island configuration (same as Fadeev equilibrium). Throumoulopoulos et. al. \cite{Throumoulopoulos2009} have constructed a class of magnetic island equilibrium in presence of shear flows by solving the Grad-Shafranov equation. Considering only sub-\Alfvenic shear flows that are relevant to fusion experiments, they have shown stabilization of island equilibrium and formation of pressure islands. Analytical work by Waelbroeck et. al. \cite{Waelbroeck2007}, using a resistive MHD model, verifies modest influence on the stability of magnetic islands when subjected to a sub-\Alfvenic shear flow. The aim of our present work is to understand the effect of shear flow on the island coalescence problem for a wide range of values of flow amplitude ($v_0$) from sub-Alfvenic to super-Alfvenic and for a wide range of velocity shear scale length ($a_v$). Using the VR-RMHD model and very high resolution computer simulations, we have investigated Fadeev equilibrium in the presence of a tan-hyperbolic flow profile (shear flow is symmetric and anti-parallel on both sides of magnetic islands). We find that in line with previous studies, the sub-\Alfvenic flows have a negligible effect on coalescence instability as the time required to attain the peak value as well the magnitude of reconnection rate is close to that in the absence of shear flow. However, for super-\Alfvenic shear flows and irrespective of shear length scales (compared to magnetic island width), we show the existence of coalescence instability and suppression of the magnetohydrodynamic Kelvin-Helmholtz instability (MHD-KHI). {{In the absence of in-plane shear flow, previously reported numerical work \cite{Birn_Hesse2007} using fully compressible resistive MHD model on island coalescence problem has shown that the effect of compressibility has no role on the reconnection rate and island dynamics. Moreover, in the case of MHD-Kelvin-Helmholtz instability (MHD-KHI), the condition for the fastest growing mode remains unchanged for both incompressible and compressible models \cite{Miura1982}. Therefore to start with the simplest, nontrivial model, here we have considered an incompressible resistive MHD model (VR-RMHD model).}} We also present extensive results on its quasi-linear and non-linear fate for the entire range of parameters addressed here. 

The rest of our paper is organized as follows: the VR-RMHD model and the BOUT++ numerical framework used for its study are discussed in Section II. Then, a series of benchmark results for the resistive island coalescence instability in the absence of in-plane flow are discussed in Section III. In Section IV, the dynamics of magnetic island - flow shear system for different $v_0$ and $a_v$ values are discussed. Finally, our conclusions and potential future work have been discussed in Section V. 
\section{Numerical Setup}
\begin{figure}[!h]
		\includegraphics[scale=0.22]{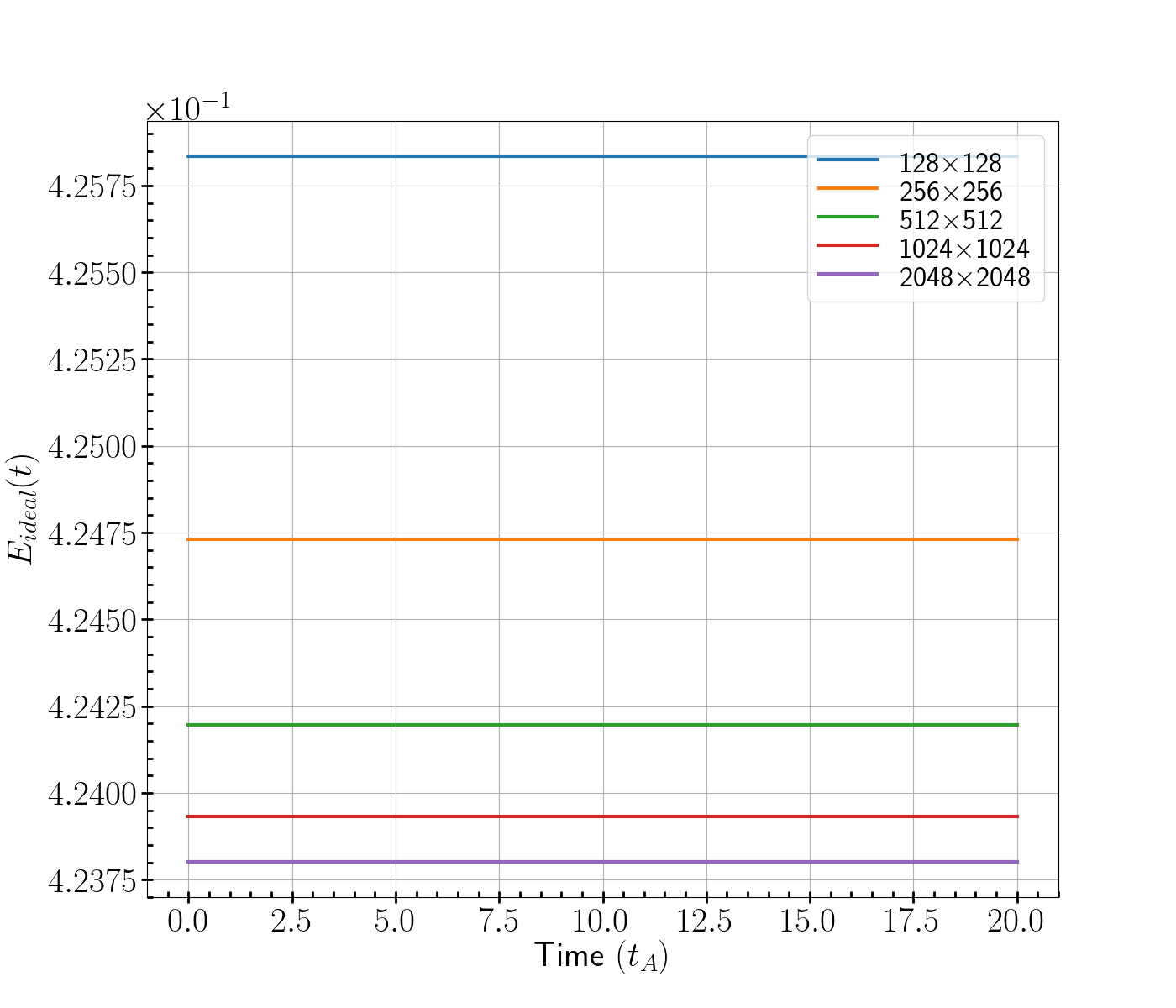}
    \caption{Total energy $E_{ideal}(t)$ in ideal MHD limit for 5 different grid sizes}. 
    \label{IMHD_E_plot}
\end{figure}

We solve the VR-RMHD model in a 2D Cartesian geometry (see Ref. \cite{rmhd1} and references there in) using the BOUT++ framework \cite{bout1,bout2}. {{As mentioned in the preceding section, this model assumes plasma as an incompressible magnetized fluid i.e. $\nabla \cdot \vel =0$, using which the full compressible MHD equations are simplified to the vorticity ($\omega$)-vector potential ($\Psi$) formalism}}. The governing equations of the $\omega$-$\Psi$ formalism read as \cite{rmhd1}
\begin{align}
&\nabla \cdot \vel = 0, \qquad \nabla \cdot \B = 0\\
&\frac{\pd \omega_y}{\pd t} = \left[\varphi,\omega_y \right]  - \left[\Psi,J_y \right] + \hat{\nu} \nabla^2 \omega_y \label{eqn:omegay-2d}\\
&\frac{\pd \Psi}{\pd t} = \left[\varphi,\Psi \right] + \hat{\eta}\nabla^2 \Psi \label{eqn:psi-2d}
\end{align}
In the above equations, the out-of-plane or $y-$component of vorticity $\omega=\hat{y} \cdot \left(\vec{\nabla} \times \vel \right) = -\nabla_{\perp}^2 \varphi$; $u_x=-\pd_z \varphi$, $u_z=\pd_x \varphi$, where $\vel ~(=u_x \hat{x} + u_z \hat{z})$ is the in-plane (``poloidal") velocity  of the plasma and $\varphi$ is the corresponding stream function. Similarly, the out-of-plane current density $J_y =\hat{y} \cdot \left( \vec{\nabla} \times \B \right)= -\nabla_{\perp}^2 \Psi$; $B_x=-\pd_z \Psi$, $B_z=\pd_x \Psi$, where $\B~(=B_x \hat{x} + B_z \hat{z})$ is the in-plane magnetic field and $\Psi$ is the corresponding magnetic vector potential. The above equations are normalized as follows: length $L$ to system length $L_x$, velocity to \Alfven velocity $v_A = B/\sqrt{\mu_0 \rho}$ and time to the \Alfvenic time $t_A = L /v_A$. {{Here, normalized density of plasma $\rho=1$ and magnetic permeability $\mu_0=1$. The normalized viscosity $\hat{\nu}$ and resistivity $\hat{\eta}$ are defined as $\hat{\nu} = \nu / (L v_A)$ and $\hat{\eta} = \eta / (\mu_0 L v_A)$, where $\nu$ is the kinematic viscosity and $\eta$ is a constant resistivity. The main variables $\omega_y$ and $\Psi$ are normalized as $\omega_y L/v_A$ and $\Psi/(B L)$}}. The Poisson bracket is defined as $\left[f,g\right]_{z,x}= (\pd_z f) (\pd_x g) - (\pd_x f) (\pd_z g)$. \\
\indent
Equations (\ref{eqn:omegay-2d}) and (\ref{eqn:psi-2d}) are solved in a 2D uniform Cartesian grid ($0\leq x \leq L_x$ and $0 \leq z \leq L_z$) where $L_x=L_z$. {{For the rest of the paper, we use dimensionless, normalized quantities unless otherwise specified.}}  Initial equilibrium current density $J_{y0}$, vorticity $\omega_{y0}$ and perturbed vector potential $\Psi_1$ profiles are\cite{rmhd1},
\begin{equation}
\left.
\begin{aligned}
&J_{y0} = \frac{1-\epsilon^2}{ a_B \left[\cosh \left( \frac{x-L_x/2}{a_B} \right) + \epsilon \cos\left(\frac{z}{a_B} \right) \right]^2} \\
& \implies \Psi_{0} = -a_B~ \text{ln}\left[ \cosh \left( \frac{x-L_x/2}{a_B}\right) + \epsilon \cos \left( \frac{z}{a_B}\right) \right]\\
&\omega_{y0} = \frac{v_0}{a_v \left[\cosh \left(\frac{x-L_x/2}{a_v}\right)  \right]^2} \\
&\Psi_1 = A \cos(2\pi z/L_z) \sin(\pi x /L_x)
\end{aligned}
\right\} \label{eqn:initial_profil}
\end{equation}
Here, the parameter $\epsilon=0.2$ determines the island width, {{$a_B=1/2\pi$ determines the simulation domain size as $L_x=L_z=4\pi a_B = 2$, $\Psi_1$ is the perturbed vector potential with amplitude $A=0.01$. For all our simulations, we have set $\hat{\eta}=\hat{\nu}=10^{-4}$. In the limit of $\epsilon = 0$, Fadeev equilibrium reduces to the Harris current sheet of shear width $a_B$. Fig. \ref{flow_av-2aB_v0-1p4_profile_a} shows spatial profiles of $J_{y0}$ (left panel colormap) and $\Psi_0$ (left panel contour) and $\omega_{y0}$ (right panel colormap, for $a_v=2a_B$ case). Periodic and Dirichlet boundary conditions are respectively implemented along the $z$ and $x$ directions. In BOUT++, we have used FFT based calculations along $z$ direction and finite difference based calculations along $x$ direction. Value of $J_y,~ \Psi, ~ \omega_y$ and $\varphi$ at $x=$ 0 and 2 is zero (Dirichlet boundary)}}.  All the runs are carried out for two different grid sizes $\mathrm{d}x=4\times10^{-3}$, $\mathrm{d}z=2\times10^{-3}$  $(N_x=512,N_z=1024)$ and $\mathrm{d}x=2\times10^{-3}$, $\mathrm{d} z=10^{-3}$ (corresponding $N_x=1024, N_z=2048)$ 
{{As given in Refs. \cite{rmhd2,rmhd3}, the width of the reconnection current sheet (say $\delta$) generated during island coalescence $\sim$ $\hat{\eta}^{2/3}$ when $\hat{\eta} \geq 10^{-4}$}}. In our case, $\hat{\eta} = 1\times10^{-4}$ (corresponding $\delta\simeq 0.00215)$, hence our $z$-directional grid size is sufficient enough to resolve the reconnection current sheet. \\
\indent 
Equations (\ref{eqn:omegay-2d})-(\ref{eqn:psi-2d}) are solved using the BOUT++ framework {{which was originally developed for tokamak edge turbulence studies}}. The BOUT++ framework uses a finite difference technique based 3D nonlinear solver. Recent developments and the use of ``Object Oriented Programming (OOP)" concepts of C++ language have enabled the users to solve an arbitrary number of coupled, general partial differential equations \cite{bout1,bout2}. In solving our set of VR-RMHD equations, we have used an energy conserving Arakawa bracket method for calculating the nonlinear terms and the implicit CVODE time solver from the SUNDIALS \cite{CVODE} package. {{For all the runs presented here, the time solver takes a time step of 0.05$t_A$}}.\\
\indent
To check the correctness of the newly implemented model, we first test for energy conservation. {{In ideal-RMHD, total energy ($E_{ideal} = 1/2\int \mathrm{d}\tau ( |\nabla_\perp \varphi|^2 + |\nabla_\perp \Psi|^2$)) of the system remains conserved. Taking $J_{y0}$ (see Eq. \ref{eqn:initial_profil}) as the initial profile and $\hat{\eta}=\hat{\nu}=0$, $E_{ideal}(t)$ as a function of time is plotted in Fig. \ref{IMHD_E_plot}. Constant $E_{ideal}(t)$ values at our presently working grid sizes (1024$\times$1024 and 2048$\times$2048) shows energy conservation by our solver in the ideal limit. Also, this verifies negligible numerical dissipation even at large grid sizes. As can be expected, the solver numerically sustains Fadeev equilibrium. 
Here, for our 2D model, the integration is over $\mathrm{d}\tau=\mathrm{d}x \mathrm{d}z$ and the operator $\nabla_\perp=\hat{x}\partial_x + \hat{z}\partial_z$}}. 
One of the important parameters to investigate is the reconnection rate. In literature \cite{rmhd1}, this reconnection rate ($M$) is calculated from the reconnection flux $\psi_r$ as $M=\pd_t \psi_r$ {{($\psi_r = \Psi_X - \Psi_O$, with $\Psi_X$ and $\Psi_O$ the magnetic flux at $X$-point and O-point, respectively)}}. {{In the absence of in-plane shear flows, the line joining the O-points always aligns itself with the line $x=1$ (see for instance Fig. \ref{no_flow_island_evolution}), hence it becomes easy to calculate $M$ from the time derivative of $\psi_r$. However, in the presence of shear flows, the line joining the O-points gets twisted because of flow dynamics and it becomes numerically expensive to compute the reconnection rate from reconnection flux. {{Reconnection rate can also be calculated by measuring the reconnection electric field $E_y$ at the $X$-point using Eq. \ref{eqn:psi-2d}, as $E_y = -\eta J_y |_{\scriptscriptstyle X}$ \cite{rmhd1,Stainner2013}. In the absence of shear flows, the nonlinear term (first term in right side of Eq. \ref{eqn:psi-2d}) has negligible contribution for the calculation of $E_y$ inside the reconnecting current sheet \cite{rmhd1}. In the presence of shear flows, this is also found to be valid}}. Hence, in this work, the reconnection rate is calculated as $M=E_y=-\eta J_y |_{\scriptscriptstyle X}$ for our benchmark studies in the absence of shear flows as well as studies with shear flows}}.
\section{Island coalescence without shear flows: benchmark}
For the benchmark purpose, we have considered the resistive island coalescence problem \cite{rmhd1,rmhd2,rmhd3} without any shear flow. {{Benchmark study uses the initial profile of equilibrium $J_{y0}$ and perturbed $\Psi_1$ (as given by Eq. \ref{eqn:initial_profil}) with $v_0=0$}}. We have used the grid size $\mathrm{d}z = 5\times10^{-4}$, $\mathrm{d}x=1\times10^{-3}$. Six different resistivity values have been considered between $5\times10^{-6}$ and $2\times10^{-4}$ (see Fig. \ref{rec-rate_eta_scan}). Here our length scale $L=2$, whereas it is $L=1$ in Ref. \cite{rmhd1} because of their quarter domain simulation; our time scale is therefore twice of that in Ref. \cite{rmhd1}.
\begin{figure*}[!h]
 \begin{subfigure}{1\textwidth}
   \includegraphics[scale=0.6]{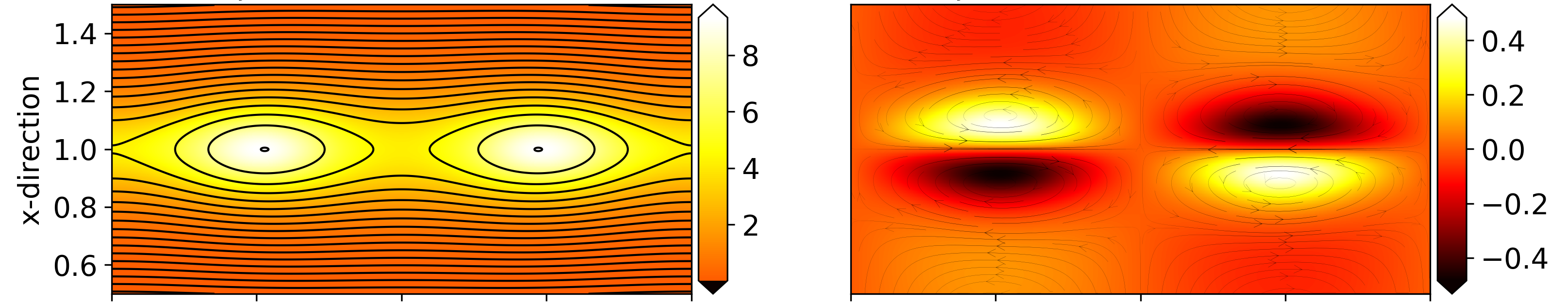}
   \caption{}
   \label{no_flow_profile_a}
  \end{subfigure}
  \begin{subfigure}{1\textwidth}
  \includegraphics[scale=0.6]{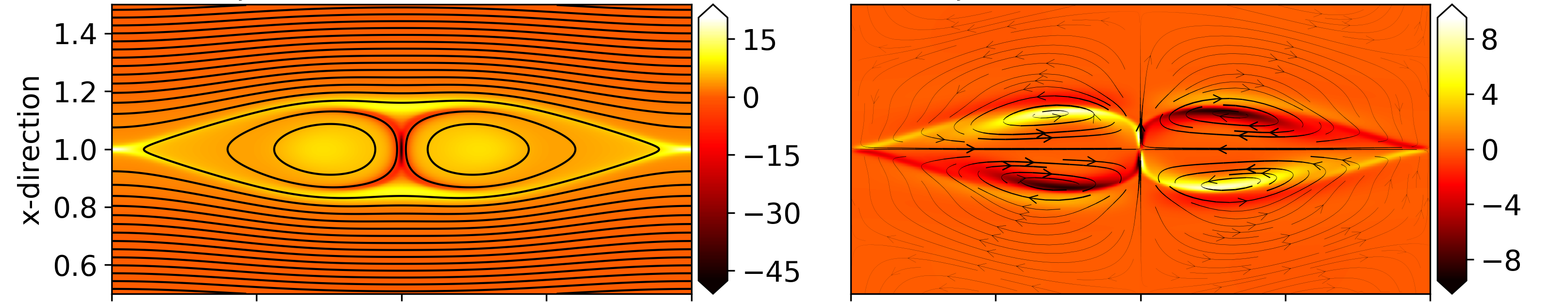}
     \caption{}
   \label{no_flow_profile_b}
  \end{subfigure}
  \begin{subfigure}{1\textwidth}
   \centering
   \includegraphics[scale=0.6]{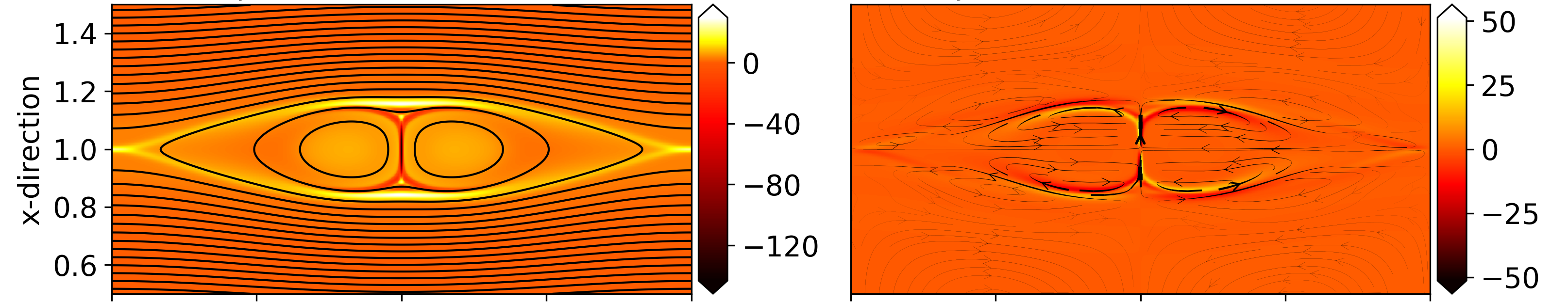}
     \caption{}
   \label{no_flow_profile_c}
  \end{subfigure}
  \begin{subfigure}{1\textwidth}
   \centering
  \includegraphics[scale=0.6]{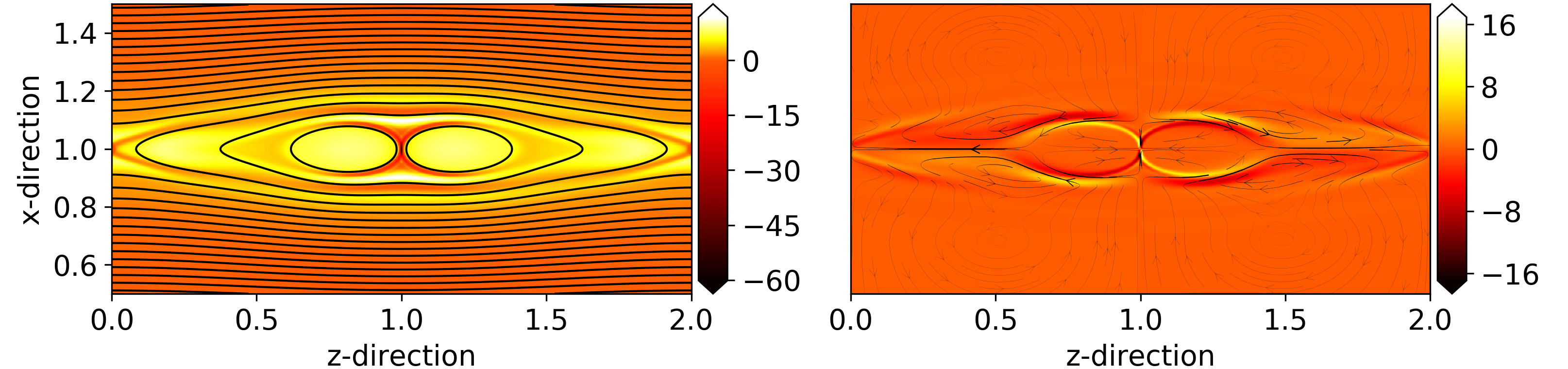}  
     \caption{}
   \label{no_flow_profile_d}
  \end{subfigure}
  \caption{Fadeev equilibrium evolution in the absence of shear flow. Left panel shows $J_y$ (colormap), $\Psi$ (contours) and right panel shows $\omega_y$ (colormap), velocity (streamlines) at various time points $t$: (a) $0.4 t_A$ (b) $2.6 t_A$ (c) $3.3 t_A$ and (d) $5.0t_A$. }
  \label{no_flow_island_evolution}
\end{figure*}
In the absence of in-plane shear flow, the time evolution of a Fadeev equilibrium is shown in Fig. \ref{no_flow_island_evolution}, similar to Fig. 2 of Ref. \cite{rmhd1}. {{As discussed earlier, Fadeev equilibrium is an ideal MHD equilibrium, hence when perturbed with finite resistivity, the plasma is able to break the frozen-in condition and cross the $X$-point \cite{finn_Kaw1977}. This allows the attraction force between the current filaments to become dominant. The typical perturbation profile $\Psi_1$, having a maximum of magnetic flux at the $X$-point, further accelerates the instability. This movement of islands towards the $X$-point is shown in Fig. \ref{no_flow_profile_a}}}. Fig. \ref{no_flow_profile_b} shows that the coalescence process has started and the reconnection sheet has formed at the $X$-point. In Fig \ref{no_flow_profile_c}, the current sheet is fully developed and the reconnection rate is at its maximum. {{After this, the magnetic flux piles up on both sides of the current sheet causing a slow down of the coalescence process, and hence the reconnection rate}}. The reconnection rate at these time frames are marked in Fig. \ref{no_flow_rec-rate_time}. In Fig. 1 of Ref \cite{rmhd1}, the reconnection rate for $\hat{\eta} = 2\times10^{-5}$ attains a peak value at $t \simeq 7 t_A$. Likewise in our case, as seen in Fig. \ref{no_flow_rec-rate_time}, $E_y$ peaks around $t \simeq 3.3 t_A$ (recall that our $t_A$ is twice that defined in Ref. \cite{rmhd1}). \\
\begin{figure}[!h]
	\begin{subfigure}{0.49\textwidth}
		\includegraphics[scale=0.26]{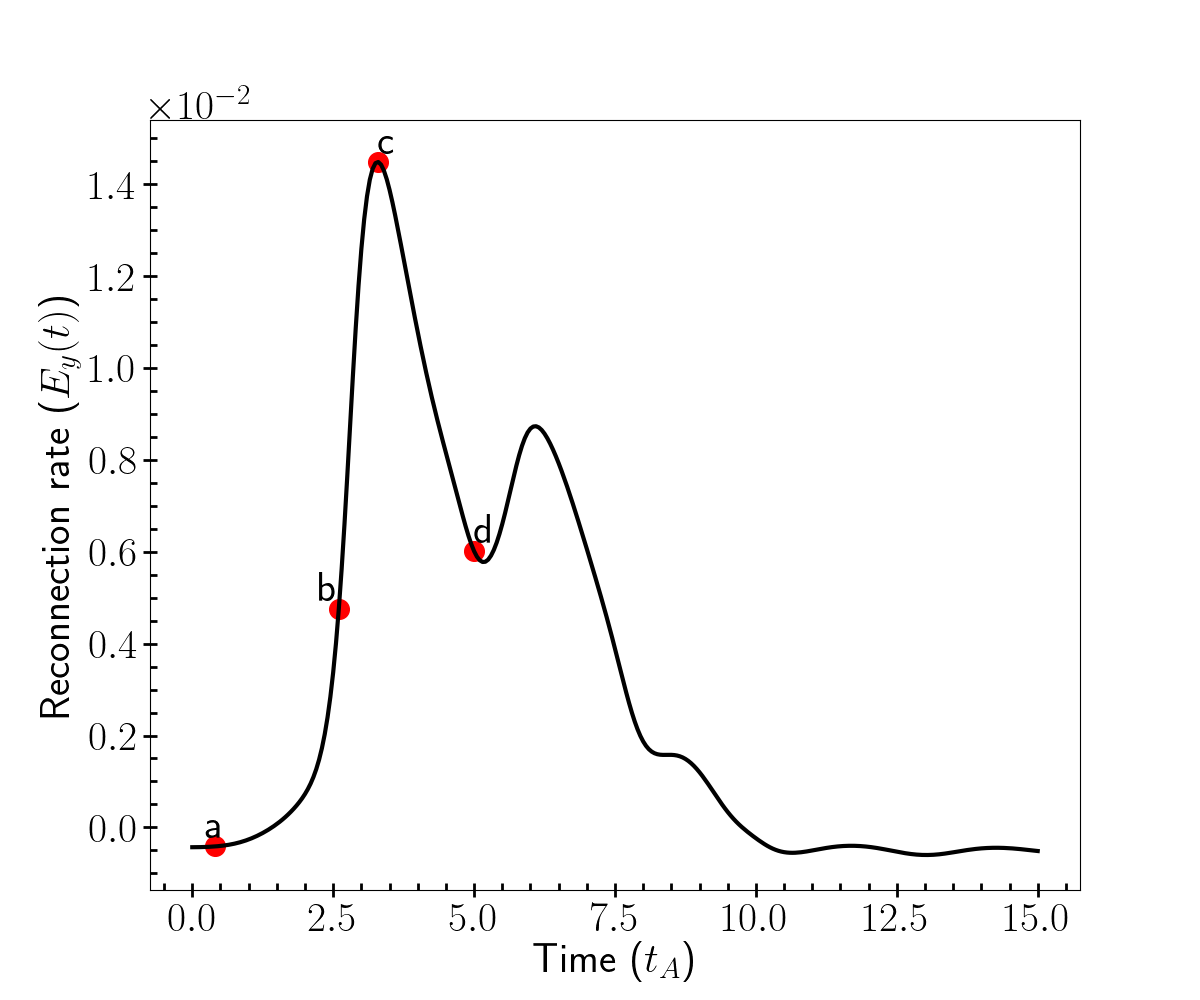}
		\caption{}
		\label{no_flow_rec-rate_time}
	\end{subfigure}
	\begin{subfigure}{0.49\textwidth}
		\includegraphics[scale=0.26]{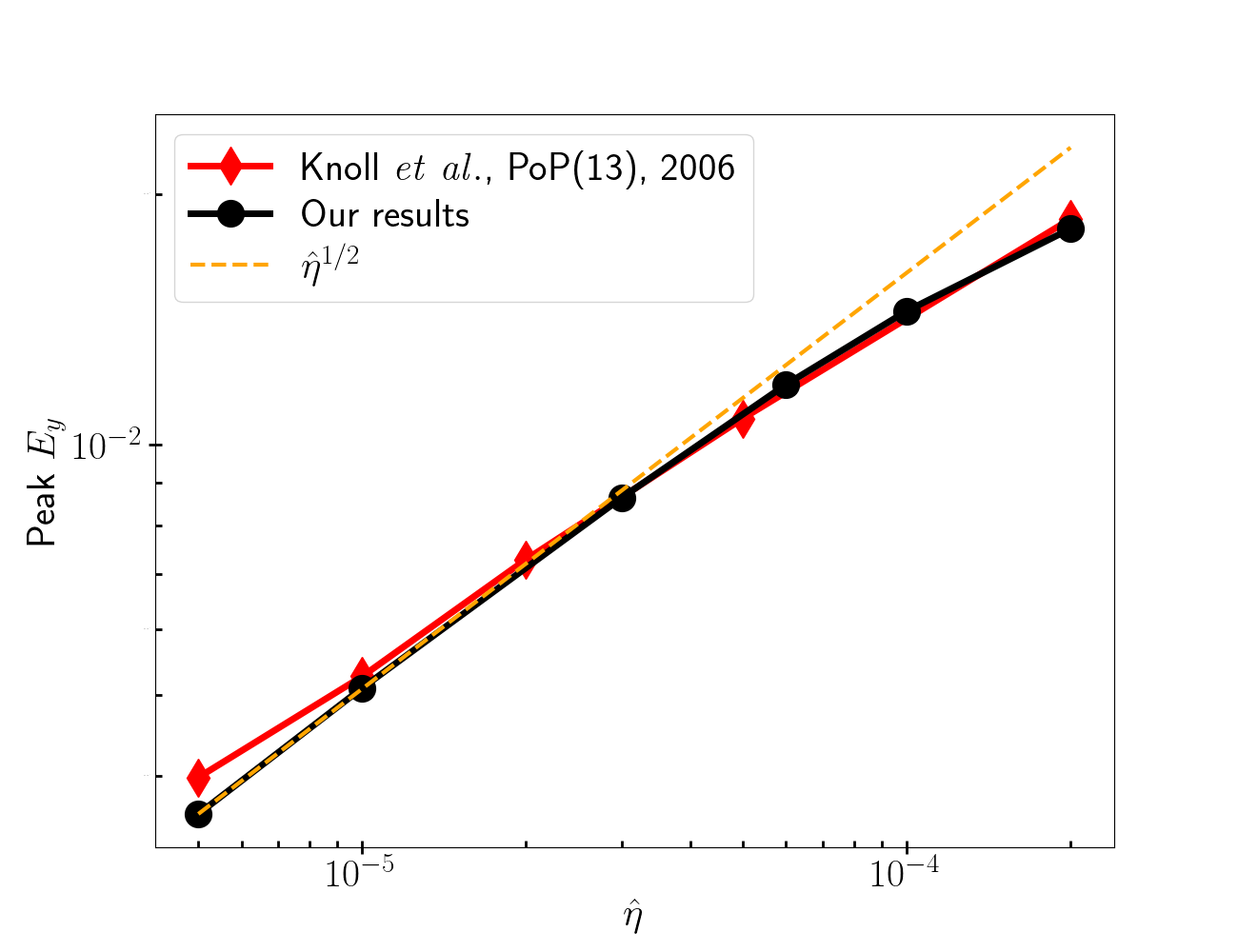}
		\caption{}
		\label{rec-rate_eta_scan}
	\end{subfigure}
	\caption{Left panel shows the reconnection rate vs. time plot for $\hat{\nu}=\hat{\eta}=10^{-4}$. Points (a), (b), (c) and (d) correspond to those shown in Fig. \ref{no_flow_island_evolution}. Right panel shows the maximum reconnection rate vs. normalized resistivity.}
\end{figure}
\indent 
Furthermore, to verify the dependence of reconnection rate on resistivity, we have plotted the maximum reconnection rate versus different $\hat{\eta}$ values in Fig. \ref{rec-rate_eta_scan}. As reported in Ref. \cite{rmhd1}, the reconnection rate scales as $\varpropto \hat{\eta}^{1/2}$ (Sweet-Parker scaling) for resistivity lower than a critical value \cite{Biskamp1986} (here, for $\hat{\eta} \leq 10^{-4}$). This clearly verifies the correctness of our solver. 
\section{Island coalescence with shear flows}
We now turn on the shear flow through an initial vorticity profile as given in Eq. \ref{eqn:initial_profil}. The shear flow profile and simulation parameters are chosen such that in the absence of current filaments ($\epsilon\rightarrow0$), the KHI gets destabilized. For our presently chosen domain size ($L_z$) and parameters ($v_0=1.4v_A$ and different values of $a_v$), we use the analytical formula given by Miura \cite{Miura1982} to calculate the fastest growing MHD-KHI mode (of mode number $m$) given as, $m = L_z/2\pi a_v$. In Table \ref{table1}, we have listed the values of $m$ for two different $L_z$ value.
\begin{table}[h]
	\caption{Mode number of fastest growing MHD-KHI mode for different $a_v$ and $L_z$ values.}
	\begin{tabular}{||c|c|c||} 
		\hline
		$\quad a_v \quad$ & \quad $m$ value for $L_z=2$ \quad & \quad $m$ value for $L_z=4$ \quad \\ [0.5ex] 
		\hline\hline
		$2a_B$ & {\bf{0.4}} & 0.8 $\sim$ {\bf{1}} \\ [0.2ex]
		\hline
		$a_B$ & 0.8 $\sim$ {\bf{1}} & 1.6 $\sim$ {\bf{2}} \\[0.2ex]
		\hline
		$a_B/2$ & 1.6 $\sim$ {\bf{2}} & 3.2 $\sim$ {\bf{3}} \\[0.2ex]
		\hline
		$a_B/4$ & 3.2 $\sim$ {\bf{3}} & 6.4 $\sim$ {\bf{6}} \\[0.2ex] 
		\hline
	\end{tabular} 
	\label{table1}
\end{table}
\begin{figure}[!h]
	\includegraphics[scale=0.35]{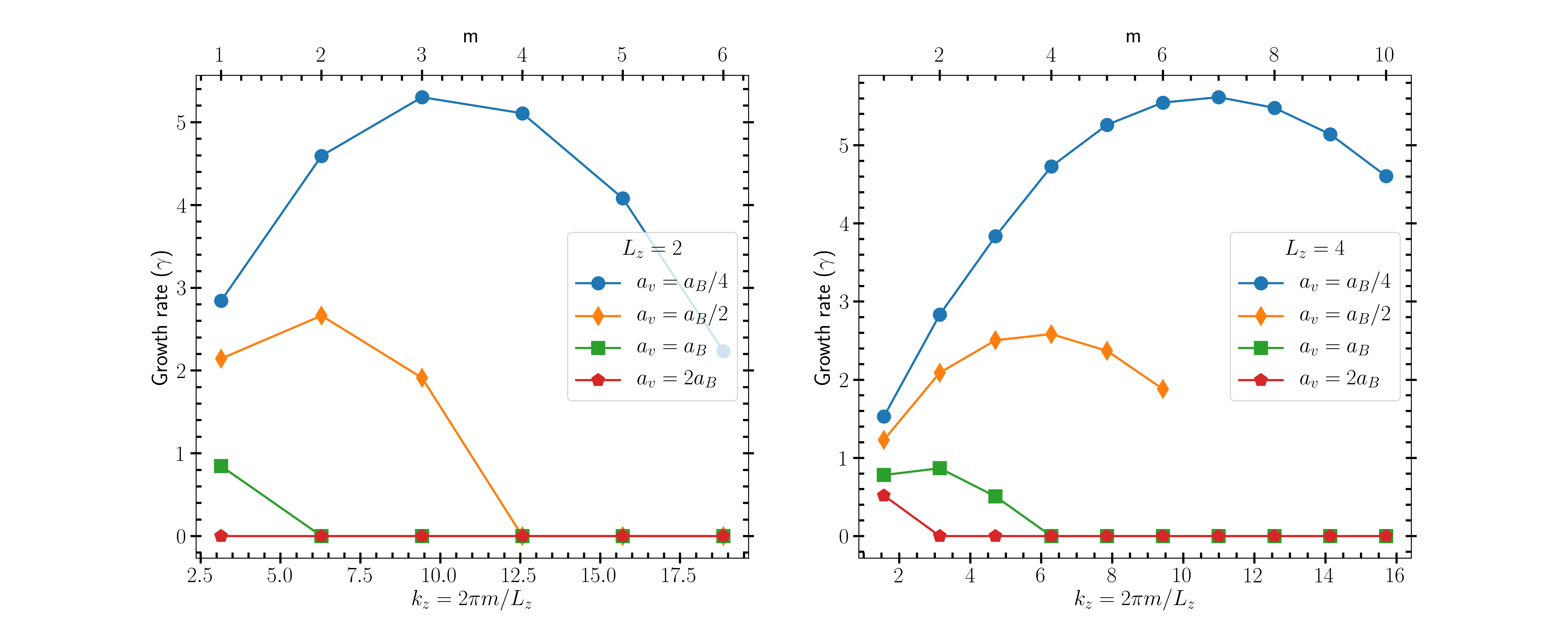}
	\caption{Growth rate of MHD-KHI mode for different wave numbers at two different system size $L_z = 2$ (left panel) and $L_z=4$ (right panel).}
	\label{gamma_vs_m}
\end{figure}
To verify the stability of these modes indicated, we perturb the  velocity shear profile with different mode numbers. Left panel of Fig. \ref{gamma_vs_m} shows the variation of growth rate for different modes for various $a_v$ values and domain sizes. As expected, we do not get an unstable mode for $a_v = 2a_B$ for domain size $L_z = 2$. {{To get an unstable mode for this velocity shear width, we double both $L_x$ and $L_z$ i.e. from $L_z=L_x=2$ to $L_z=L_x= 4$ as well as grid numbers $N_z$ and $N_x$ in order to keep the grid resolution same. As shown in the right panel of Fig. \ref{gamma_vs_m}, doubling the length $L_x$ (for $L_z=4$ case) takes the x-boundary farther from the vorticity sheet, resulting a marginal change in growth rate of the MHD-KHI modes. The fastest growing mode for $a_v=2a_B$ and $L_z=4$ is found to be $m=1$.}} \\
\indent Effects of in-plane shear flow in the island coalescence problem are studied by changing three important parameters in the vorticity (or velocity) profile: (1) flow shear strength $v_0$ (2) shear width $a_v$ and (3) the direction of shear flow parallel or anti-parallel to the magnetic field. Parameters used for these simulation are given in Section II. 
\subsection{Varying flow shear amplitude $v_0$}\label{sec: av=2aB-Lx2}
At first, we study the effect of shear flow with different amplitude values, keeping the shear length scale fixed at $a_v = 2a_B$ (shear flow length scale is larger than the island size, see Fig. \ref{shear_width_profile_full}).
\begin{figure}[!h]
  \includegraphics[scale=0.29]{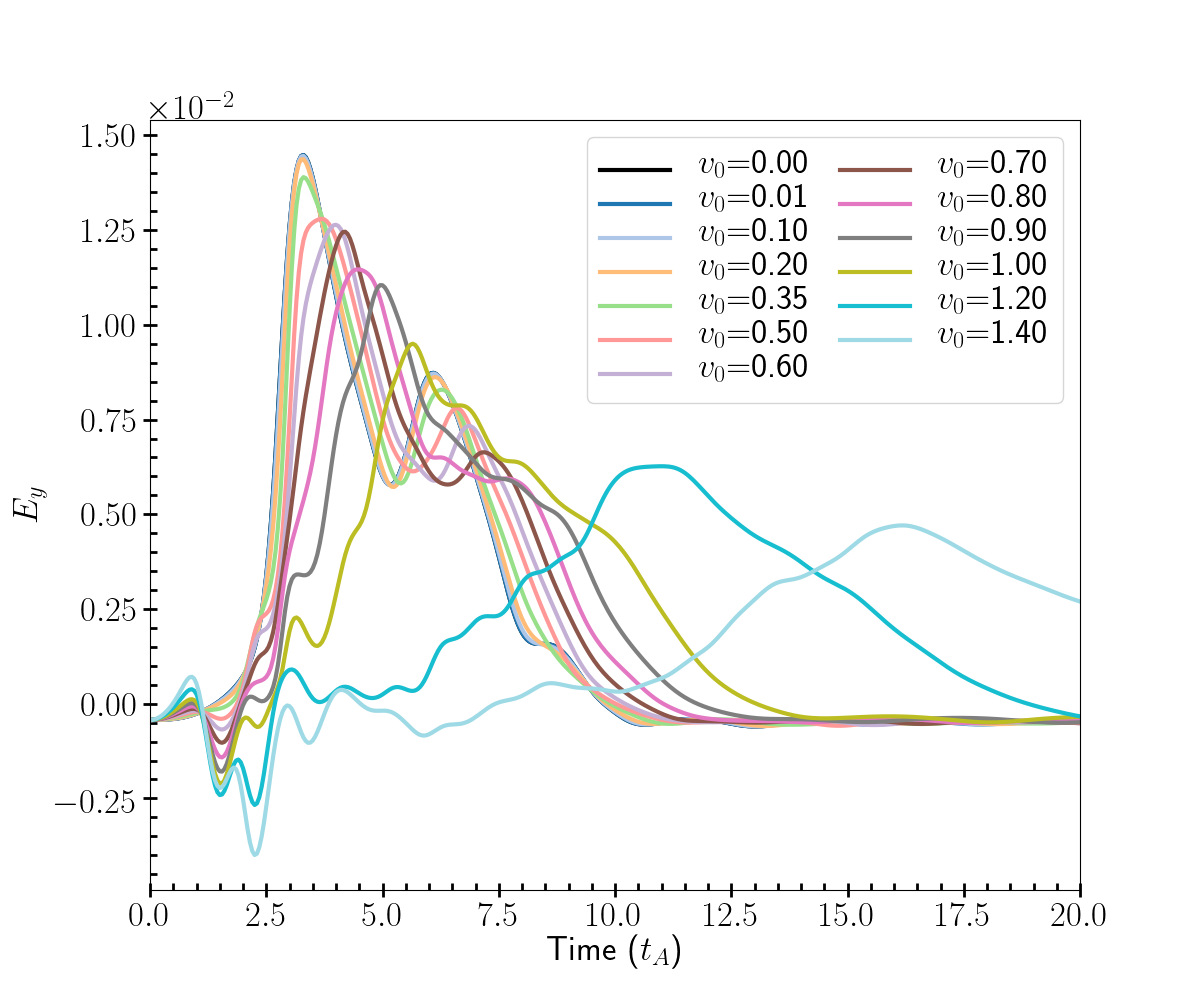}
  \caption{Time evolution of the reconnection electric field ($E_y$) at the $X$-point for $a_v=2a_B$, with $v_0$ varied between $0.01 v_A$ to $1.4v_A$.}
  \label{Ey-time-av-2aB}
\end{figure}%
\begin{figure*}[!h]
 \begin{subfigure}{1\textwidth}
   \includegraphics[scale=0.55]{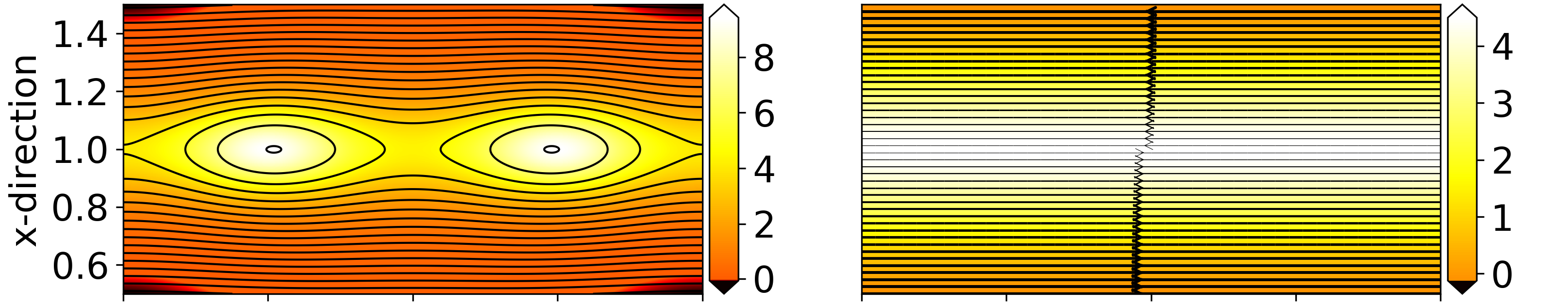}
   \caption{}
   \label{flow_av-2aB_v0-1p4_profile_a}
  \end{subfigure}
  \begin{subfigure}{1\textwidth}
  \includegraphics[scale=0.55]{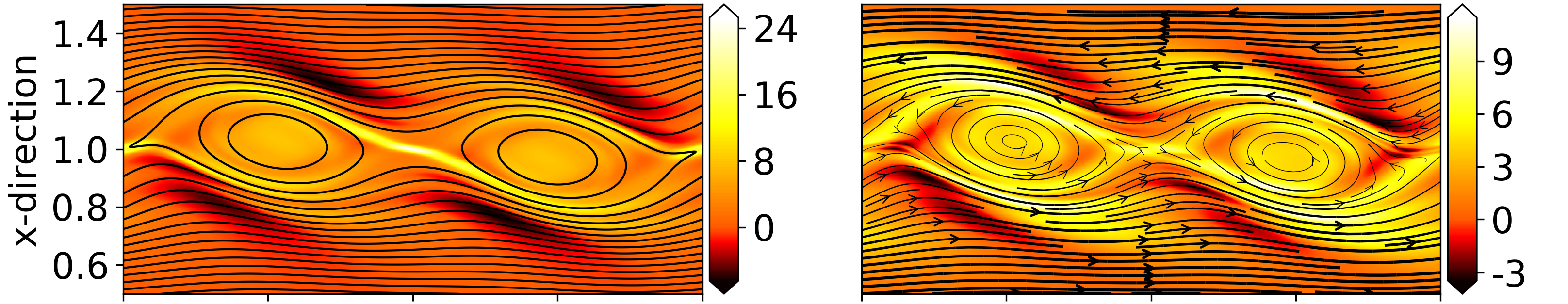}
     \caption{}
   \label{flow_av-2aB_v0-1p4_profile_b}
  \end{subfigure}
  \begin{subfigure}{1\textwidth}
   \centering
  \includegraphics[scale=0.55]{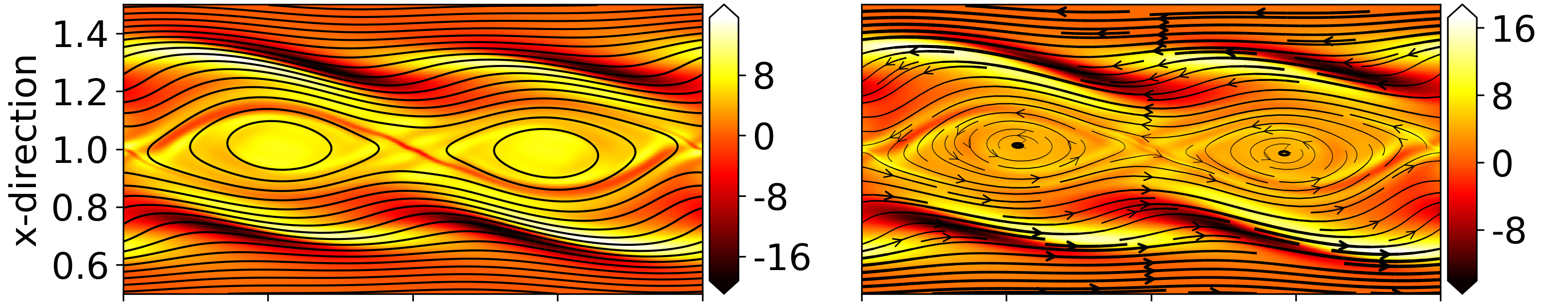}
     \caption{}
   \label{flow_av-2aB_v0-1p4_profile_c}
  \end{subfigure}
  \begin{subfigure}{1\textwidth}
   \centering
  \includegraphics[scale=0.55]{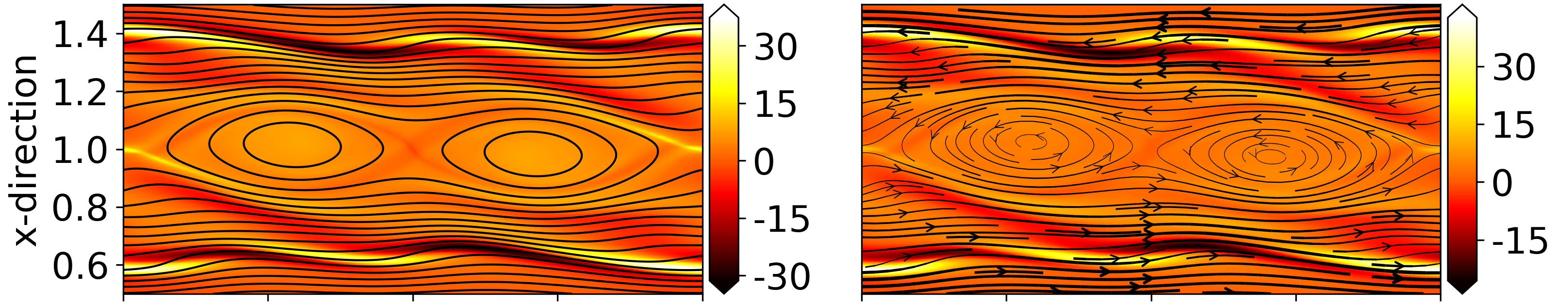}  
     \caption{}
   \label{flow_av-2aB_v0-1p4_profile_d}
  \end{subfigure}
  \begin{subfigure}{1\textwidth}
   \centering
  \includegraphics[scale=0.55]{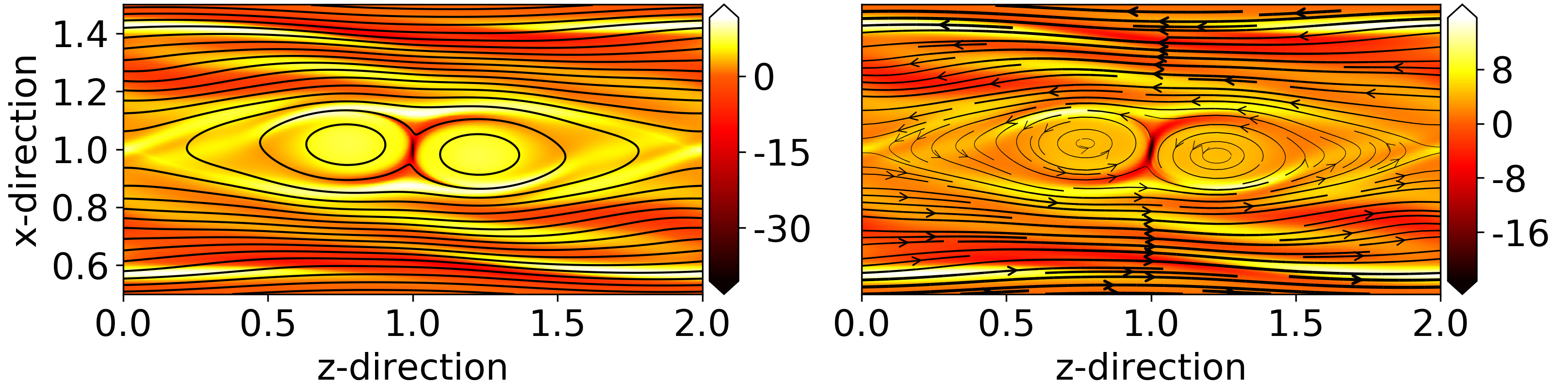}  
     \caption{}
   \label{flow_av-2aB_v0-1p4_profile_e}
  \end{subfigure}
  \caption{Left panel showing $J_y$ (colormap), $A_y$ (contours) and right panel showing $\omega$ (colormap), velocity (streamlines) at times t = (a) $0$ (b) $2t_A$ (c) $4t_A$ (d) $8t_A$ and (e) $15t_A$ for $a_v=2a_B$ and $v_0=1.4 v_A$. Width of streamlines represent flow magnitude.}
  \label{P-flow_island_evolution_av-2aB_v0-1}
\end{figure*}
{{As discussed in literature \cite{Pritchett1992,Pritchett-Wo-1979}, the $x$-directional width, $a_I$, of Fadeev equilibrium island system is decided by the parameter $\epsilon$ as $a_I=2\sqrt{\epsilon}\left(1-\epsilon/6 \right)a_B$. For $\epsilon \simeq 0.2$, $a_B \simeq a_I$ (for comparison , see Fig. \ref{shear_width_profile_full}). Here, throughout this work, the velocity shear width $a_v$ is compared with $a_B$}}. In this case study, $v_0$ is varied keeping the flow shear width fixed at $a_v=2a_B$. In Fig. \ref{Ey-time-av-2aB}, the time variation of the reconnection electric field is shown for different $v_0$ values. From this figure, it can be observed that for lower values of shear flow amplitude ($v_0  \lesssim 0.2$), the $E_y$ vs. time matches with the no-shear flow ($v_0 = 0$) curve (see Fig. \ref{no_flow_rec-rate_time}). {{For these $v_0$ values, three distinct \emph{phases} can be identified: (a) The reconnection rate first increases slowly (sub-exponential increase in $E_y$) up to the time $\simeq 2 t_A$, slowly displacing the islands from their initial positions towards the $X$-point. (b) The linear phase of the coalescence instability continues up to $\simeq 3.3 t_A$ when the peak reconnection rate is achieved; in this phase both islands accelerate towards the $X$-point (exponential increase in $E_y$) resulting in the thinning of the reconnecting current sheet. (c) This motion causes the magnetic flux to pile up on both sides of the $X$-point resulting in a slowing down of island motion and decrease in the $E_y$ value. The islands bounce back and forth several times and finally the coalescence process completes. Hence, for lower shear flow strengths (compared to $v_A$) and larger shear scale length (compared to $a_B$), in-plane flows negligibly affect the overall coalescence process.}} \\
\indent
%
For $v_0 \geq 0.35 v_A$, the slowly growing phase of $E_y$ starts showing oscillations driven by stronger shear flows trying to peel off the islands and altering the magnetic field profile in the vicinity of the $X$-point. In this phase, as $v_0$ is increased, the magnitude of oscillation in $E_y$ increases. In Fig. \ref{Ey-time-av-2aB}, one can observe that even when the strength of shear flow is super-\Alfvenic ($v_0=1.2,~1.4$), after the initial oscillations in the \emph{first phase}, the value of $E_y$ continues to increase in the \emph{second phase}. This indicates the survival as well as the continuation of current island coalescence in super-\Alfvenic shear flows when the initial shear flow scale length $a_v = 2 a_B$. Another interesting point to notice is the decrease in magnetic flux pile-up as shear flow amplitude increases. up to $v_0=0.7v_A$, one can notice bouncing of islands after initial merging (decrease in $E_y$ value after peak reconnection, see Fig \ref{Ey-time-av-2aB}). As $v_0$ increases further, there is no clear sign of the second peak in $E_y$ following the first maxima with the $E_y$ decreasing continuously. This indicates that with strong shear flows, the rate of flux pile up reduces, causing a slowing down of merging. \\
\indent
In Fig. \ref{P-flow_island_evolution_av-2aB_v0-1}, the time evolution of islands is shown for $v_0=1.4v_A$. Fig. \ref{flow_av-2aB_v0-1p4_profile_a} shows the initial profiles. The effect of shear flow can be seen in Fig. \ref{flow_av-2aB_v0-1p4_profile_b} where the islands get displaced along the $x$-direction. Also, the vorticity profile has now changed from a single sheet to an $m = 2$ ($m$ is the poloidal/z-directional mode number) like profile, similar to the initial $J_y$ profile. The velocity streamline shows plasma circulation inside the current islands. The flow speed inside the current island is much less than the outer shear flow magnitude. 
These rotational flow inside the islands (see Fig. \ref{flow_av-2aB_v0-1p4_profile_b}-\ref{flow_av-2aB_v0-1p4_profile_c}) stabilizes them against the shear flow. 
These stabilized magnetic islands eventually become susceptible to coalescence instability. {{The plasma circulation inside the islands sets up a shear flow on both sides of the current sheet, which in turn suppresses the upstream flow (flow into the reconnection sheet), causing a smaller reconnection rate. Reduced upstream flow is also responsible for less pile-up of magnetic flux. After island merging, a large island survives with rotation of plasma column.}}  As discussed in the previous section, there is no unstable MHD-KHI mode for these parameters ($L_z=2$ and $a_v=2a_B$). Hence, we found that the shear flow is unable to change the shape of current filaments and overall coalescence dynamics, but the reconnection rate decreased by $\sim$70\% (see Fig. \ref{Ey-scan-full}) when $v_0$ increases from 0 to $1.4v_A$. To see the effect of an unstable MHD-KHI mode on the coalescence process, we changed the shear width (keeping $L_z=2$) in the next subsection \ref{Sec.a_v=a_B/4} and system size ($L_z=L_x=4$ for $a_v=2a_B$ case) in the subsection \ref{Sec.Lz=4}.
\subsection{Varying velocity shear scale length $a_v$}\label{Sec.a_v=a_B/4}
To see the effect of velocity shear length, we take $a_v=2a_B, ~a_B, ~ a_B/2, ~a_B/4$, and for each $a_v$ value, $v_0$ is scanned over a range of values between $0.1v_A$ and $1.4v_A$.
\begin{figure}[!h]
\begin{subfigure}{0.49\textwidth}
   \centering
\includegraphics[scale=0.28]{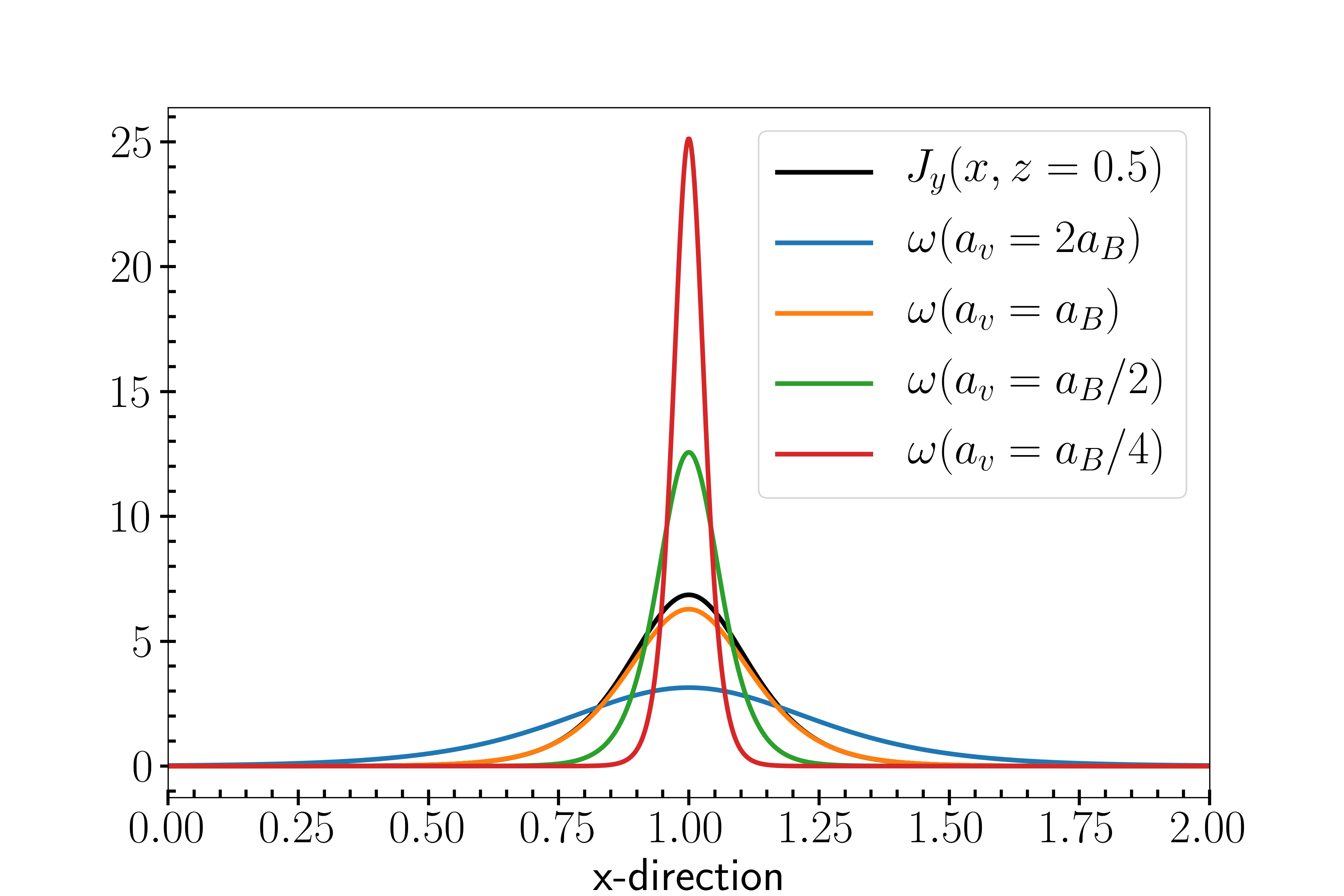}
\caption{}
\label{shear_width_profile_full}
\end{subfigure}
\begin{subfigure}{0.49\textwidth}
   \centering
\includegraphics[scale=0.28]{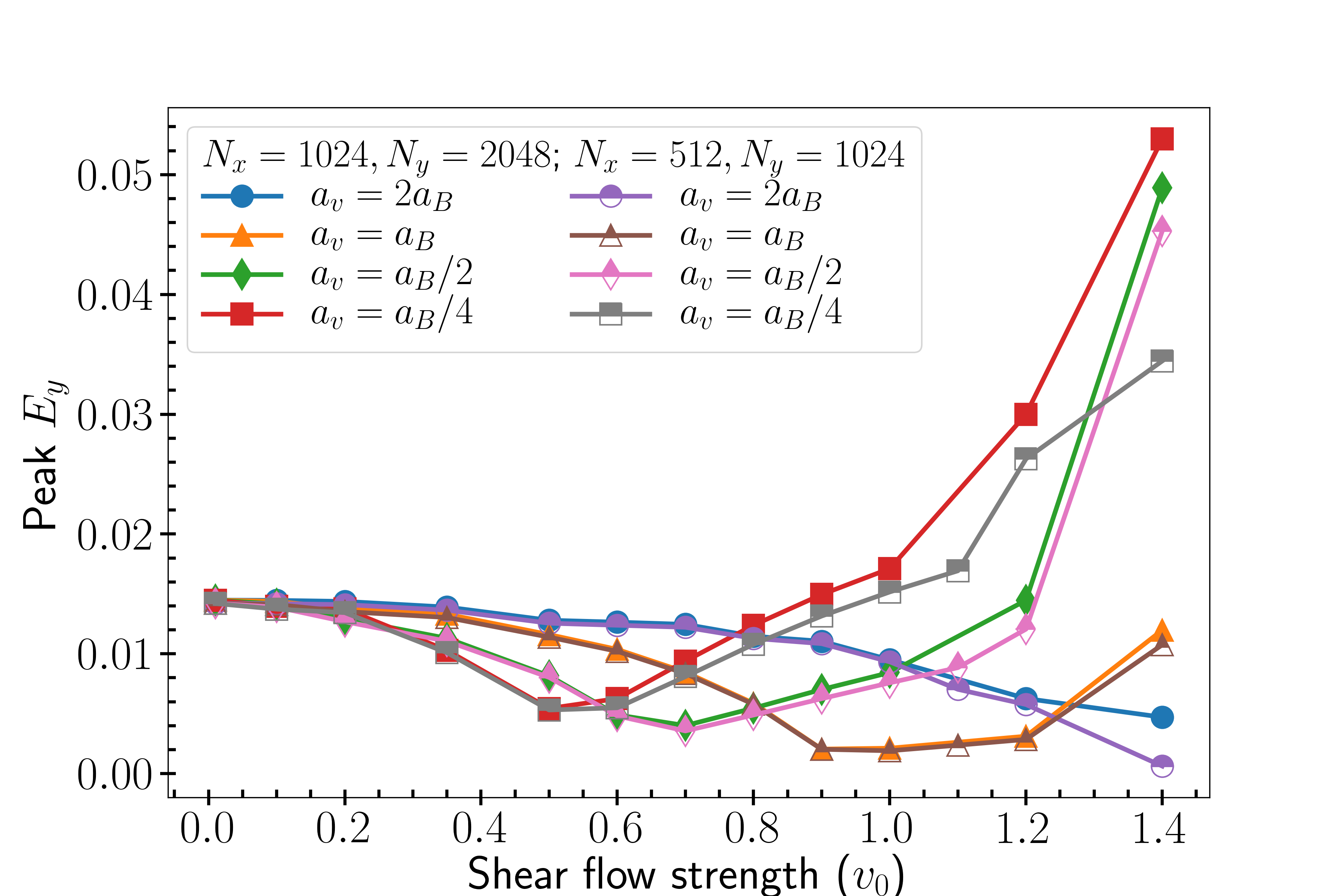}
\caption{}
\label{Ey-scan-full}
\end{subfigure}
\caption{(a) Equilibrium vorticity profiles and $J_{y0}$ profile at $z=0.5$. (b) Reconnection electric field $E_y$ vs. shear flow strength for different shear flow scale length for two set of grid sizes.}
\end{figure}
For comparison, we have plotted the shear width of the initial vorticity profiles (see Eq. \ref{eqn:initial_profil}) along with the initial $J_{y0}(x,z=0.5)$ profile at the location where the width of the island is maximum. In Fig. \ref{shear_width_profile_full}, both the velocity shear width and island width almost matches for $a_v=a_B$ (since island width $a_I \simeq a_B$). One may expect a strong influence of shear flow on the island dynamics for shear width smaller than island width. In this section, we have discussed the evolution of islands in presence of shear flow when $a_v=0.25a_B$. \\
\indent
Fig. \ref{Ey-scan-full} shows the variation in the peak values of $E_y$ for the full parametric scan over $v_0$ and $a_v$. For $a_v = 2a_B$, the peak $E_y$ value decreases monotonically with the increase of $v_0$, as discussed in the previous section. As $E_y \varpropto -J_y$, $E_y$ attains its maximum value when the current density in the reconnection current sheet is minimum (negatively maximum). As $a_v =2a_B$, current islands undergo coalescence process without much distortion. Higher $v_0$ values induce stronger plasma circulation inside the islands and this in-turn decreases the upstream velocity of plasma into the reconnection sheet causing a monotonic decrease in peak $E_y$ value. However, for $a_v \leq a_B$, shear flow is trying to destabilize the islands. For $a_v = a_B, 0.5a_B ~\text{and}~ 0.25a_B$, the  peak $E_y$ first decreases up to a critical value of $v_0$, we call it $v_{0c}$; these values are 0.9$v_A$, 0.7$v_A$ and 0.5$v_A$ respectively. Up to $v_{0c}$, the shear flow is not strong enough and in these cases, the peak $E_y$ is the reconnection rate driven by the coalescence process. For $v_0 > v_{0c}$, the shear flow becomes stronger and tries to peel off the islands. This peeling also changes the current distribution near the $X$-point, generating very thin current sheets which we are measuring as oscillations in the temporal evolution of $E_y$. Hence, here the peak $E_y$ value is not because of coalescence driven reconnection, although the islands coalesce after a long time (see Fig. \ref{P-flow_island_evolution_av-0p25aB_v0-1}). In Fig. \ref{flow_av-0p25aB_v0-1p4_profile_a}, \ref{flow_av-0p25aB_v0-1p4_profile_b} and \ref{flow_av-0p25aB_v0-1p4_profile_c}, one can clearly notice the peeling off effect of shear flow on the magnetic islands. However in Fig. \ref{flow_av-0p25aB_v0-1p4_profile_d}, at $16t_A$, vorticity patches have been formed with flow circulation coinciding with the magnetic islands. This confirms the stabilizing effect of circulation on the islands. In Fig. \ref{flow_av-0p25aB_v0-1p4_profile_e}, at $37t_A$, these surviving islands coalesce to form a large single island, as in the case of $a_v > a_B$.\\ 
\indent Comparison of peak $E_y$ vs. $v_0$ in Fig. \ref{Ey-scan-full} at two different grid size is also plotted. For $v_0 \leq v_{0c}$, the peak $E_y$ is same for both lower and higher resolution. This implies, for these cases, the current sheet at the $X$-point is well resolved by lower and higher resolution. However, for $v_0 > v_{0c}$, the shear flow generates very thin current sheets by destabilizing current islands. The lower resolution is not enough to resolve these thin current sheets. This explains the data points for lower and higher resolution are matching for $v_0 \leq v_{0c}$ but not matching for $v_0 > v_{0c}$.\\
\begin{figure*}[!h]
 \begin{subfigure}{1\textwidth}
   \includegraphics[scale=0.55]{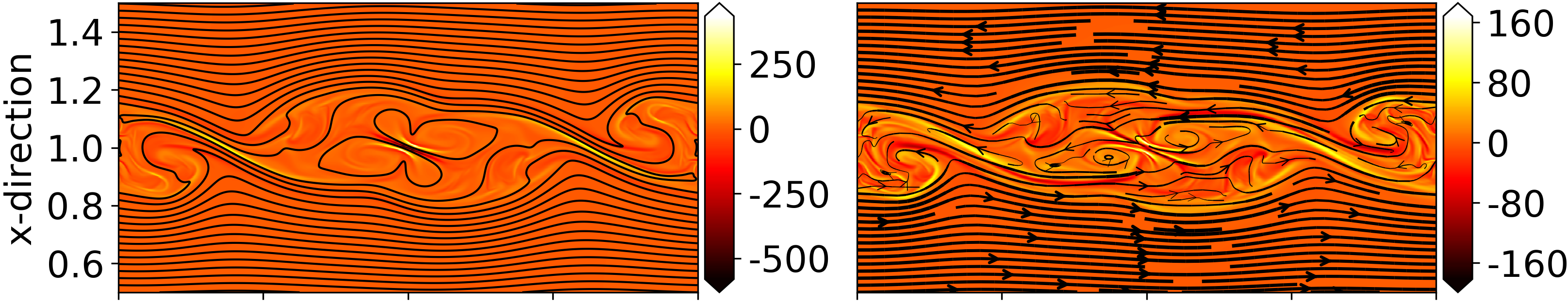}
   \caption{}
   \label{flow_av-0p25aB_v0-1p4_profile_a}
  \end{subfigure}
  \begin{subfigure}{1\textwidth}
  \includegraphics[scale=0.55]{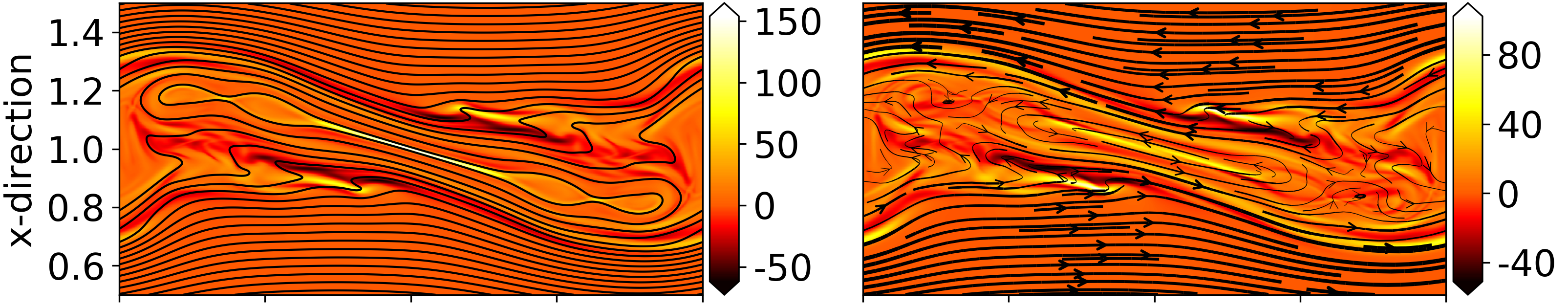}
     \caption{}
   \label{flow_av-0p25aB_v0-1p4_profile_b}
  \end{subfigure}
  \begin{subfigure}{1\textwidth}
   \centering
  \includegraphics[scale=0.55]{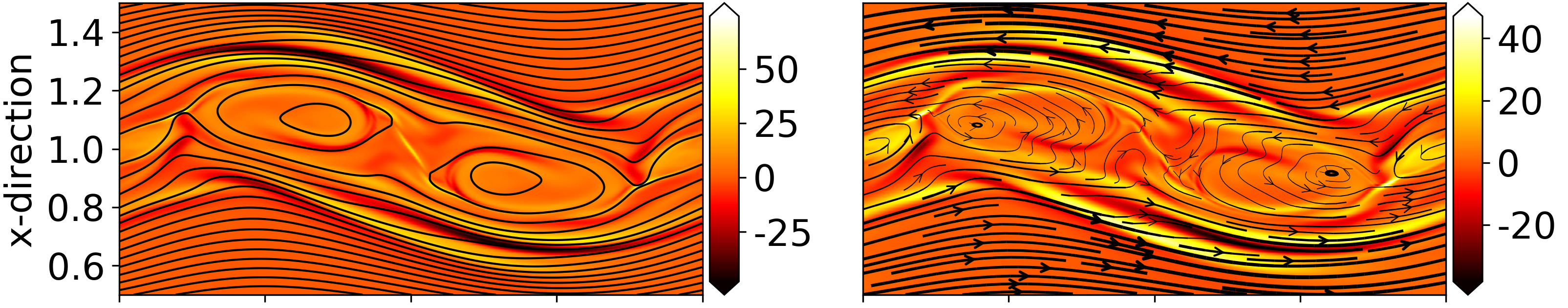}
     \caption{}
   \label{flow_av-0p25aB_v0-1p4_profile_c}
  \end{subfigure}
  \begin{subfigure}{1\textwidth}
   \centering
  \includegraphics[scale=0.55]{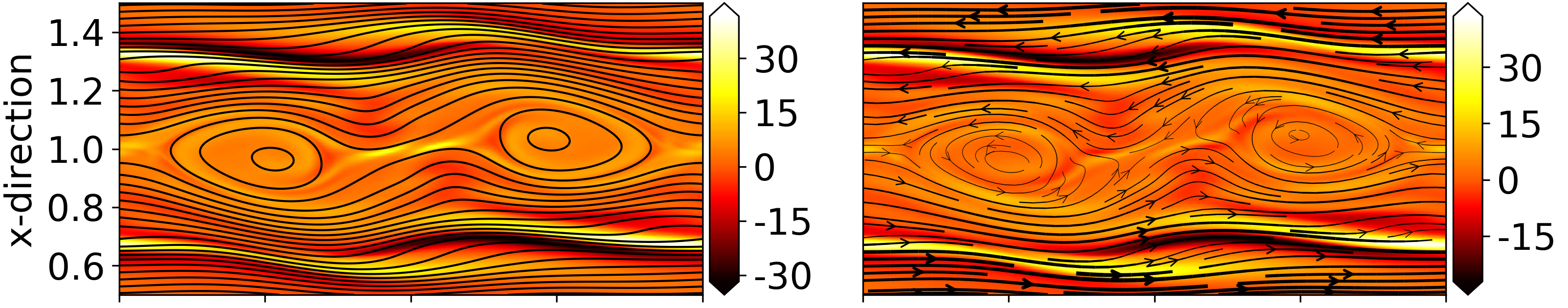}  
     \caption{}
   \label{flow_av-0p25aB_v0-1p4_profile_d}
  \end{subfigure}
  \begin{subfigure}{1\textwidth}
   \centering
  \includegraphics[scale=0.55]{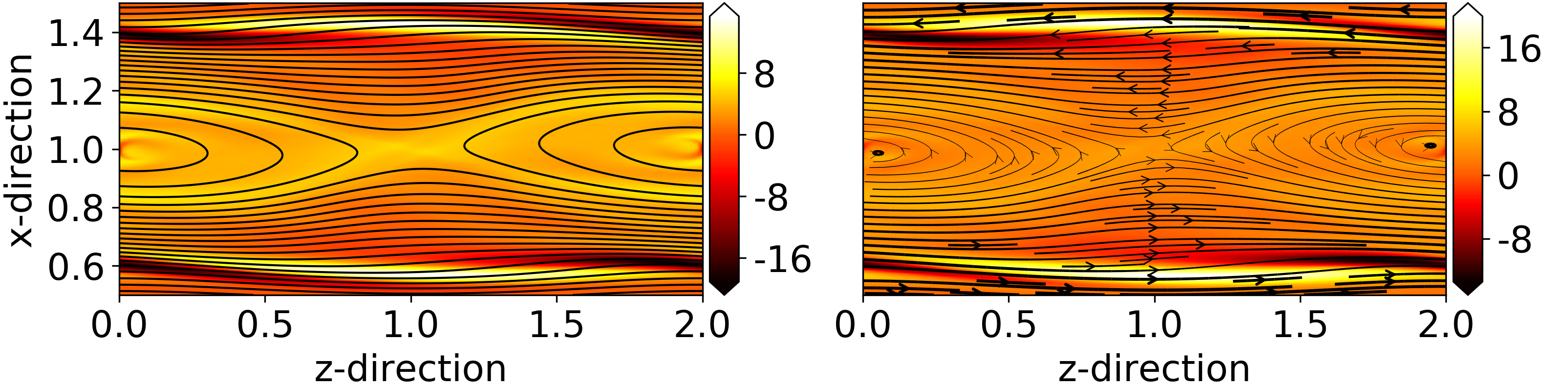}  
     \caption{}
   \label{flow_av-0p25aB_v0-1p4_profile_e}
  \end{subfigure}
  \caption{Left panel shows $J_y$ (colormap), $A_y$ (contours) and right panel shows $\omega$ (colormap), velocity (streamlines) at time t = (a) $2.25t_A$ (b) $4.5t_A$ (c) $10.0t_A$ (d) $16.0t_A$ and (e) $37.0t_A$ for $a_v=a_B/4$ and $v_0=1.4 v_A$.}
  \label{P-flow_island_evolution_av-0p25aB_v0-1}
\end{figure*}
%
%
\indent
As the MHD-KHI gets destabilized in anti-parallel magnetic field configuration \cite{Keppens1999}, shear flow anti-parallel to the magnetic field is also tested for Fadeev equilibrium. As reported in the literature for TMI case \cite{Chen1997}, we found no difference in the results compared to the parallel configuration discussed above. One explanation for this could be that the stabilizing role of the flow-induced plasma circulation inside the islands is independent of the direction of shear flow and the KHI gets suppressed for the range of $v_0$ and $a_v$ discussed here. Higher values of $v_0$ may destabilize the current island and generate KHI. Then one can observe the difference in the growth rate of KHI in parallel and anti-parallel configuration.
\subsection{Effect of varying system size $L_z$ and $L_x$}\label{Sec.Lz=4}
{{As discussed in subsection \ref{sec: av=2aB-Lx2}, there is no unstable MHD-KHI mode for $a_v=2a_B$ and $L_z=L_x=2$ domain size. Hence, for $a_v=2a_B$ case, we increase the domain size from $L_z=L_x=2$ to $L_z=L_x=4$. The magnetic island system with shear flows are perturbed with two different mode number $m=1$ (wavelength = $L_z = 4$) and $m=2$ (wavelength = $L_z/2 = 2$). In Fig. \ref{gamma_vs_m}, as mentioned earlier, growth rates of MHD-KHI modes have changed marginally when x-boundaries are taken farther. Figure \ref{island} shows the time evolution of magnetic islands in the presence of super-\Alfvenic shear flows ($v_0=1.4v_A$ and $a_v = 2a_B$) when perturbed with $m=1$ mode. At the initial times (Fig. \ref{island_a}), the strong shear flow prevents the islands to coalesce. However, as seen in Fig. \ref{island_b}, \ref{island_c}, \ref{island_d} and \ref{island_e}, the islands start to coalesce in the later time. Moreover, due to the unstable MHD-KHI mode, the coalescing point is not stationary at any particular x- or z-location, rather moves dynamically inside the blue color box marked between the z-location 0.7 - 1.0 (see Fig. \ref{island_b}, \ref{island_c} and \ref{island_d}). Hence, for this case, reconnection rate is calculated as maximum value of $E_y(x,z,t)(=-\eta J_y(x,z,t))$ inside the blue box. In Fig. \ref{Ey_plot}, one can observe a significant change in the time variation of $E_y$ and hence the island dynamics, when perturbation mode number changed from $m=2$ to $m=1$. With $m=2$ perturbation, the MHD-KHI mode is stable for both $L_x=2$ and $4$ case. Hence, the island dynamics and $E_y$ are found to be almost similar for both $L_x$ values and the coalescing point (or the reconnection point) remains stationary at z=1 and 3 location. However, with $m=1$ perturbation, the corresponding MHD-KHI mode becomes non-linearly unstable causing the island dynamics and time variation of $E_y$ is different for $L_x=2$ and $L_x=4$ case.
\begin{figure}[!h]
	\begin{subfigure}{1\textwidth}
		\includegraphics[scale=0.43]{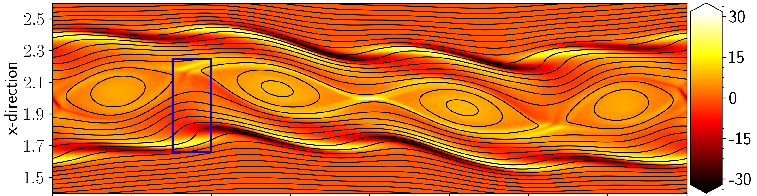}
		\caption{}
		\label{island_a}
	\end{subfigure}
	\begin{subfigure}{1\textwidth}
		\includegraphics[scale=0.43]{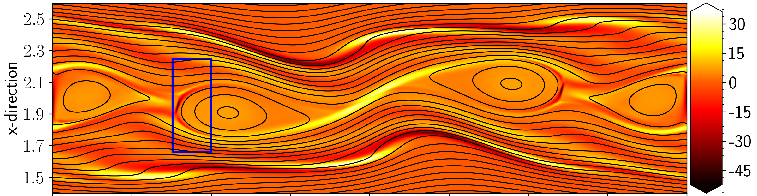}
		\caption{}
		\label{island_b}
	\end{subfigure}
	\begin{subfigure}{1\textwidth}
		\centering
		\includegraphics[scale=0.43]{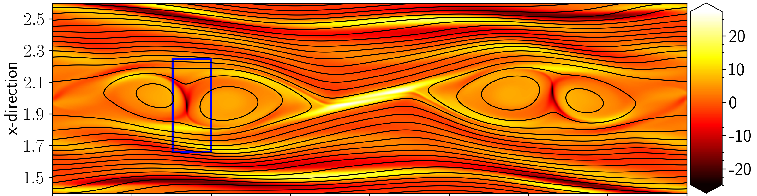}
		\caption{}
		\label{island_c}
	\end{subfigure}
	\begin{subfigure}{1\textwidth}
		\centering
		\includegraphics[scale=0.43]{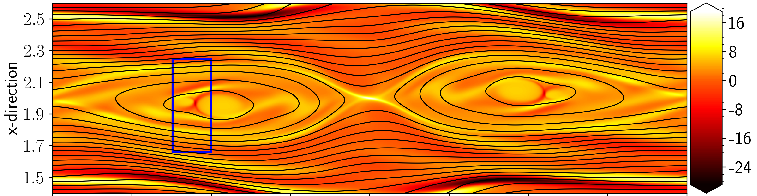}  
		\caption{}
		\label{island_d}
	\end{subfigure}
	\begin{subfigure}{1\textwidth}
		\centering
		\includegraphics[scale=0.43]{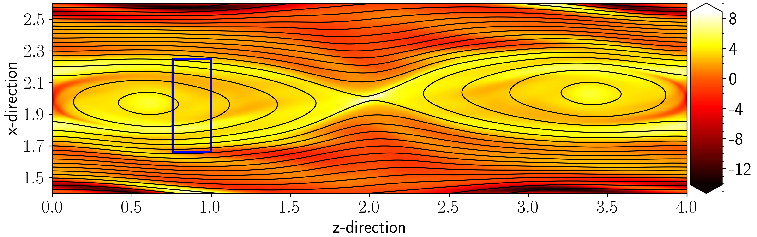}  
		\caption{}
		\label{island_e}
	\end{subfigure}
	\caption{Time evolution of $J_y$ (colormap) and $A_y$ (contour) at time: (a) 5.5$t_A$, (b) 10$t_A$, (c) 21.5$t_A$, (d) 28.5$t_A$, and (e) 45$t_A$ for shear flow parameters $a_v=2a_B$, $v_0=1.4v_A$, and domain size $L_x=L_z=4$ when perturbed with $m=1$ mode. The coalescing point moves dynamically in a region bounded by the blue-colored box.}
	\label{island}
\end{figure}
\begin{figure}[!h]
		\includegraphics[scale=0.3]{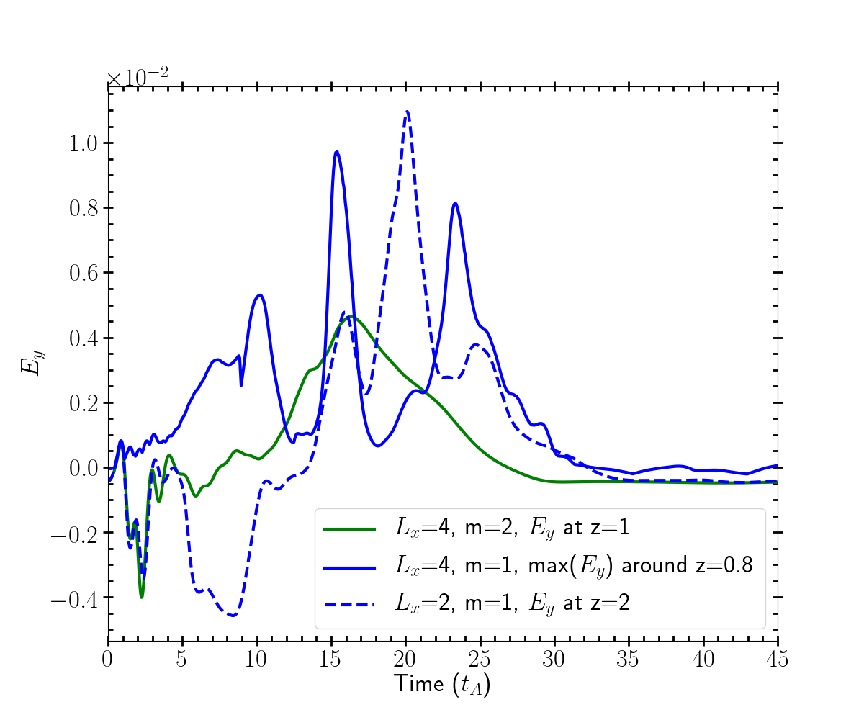}
	\caption{ Reconnecting electric field $E_y$ (or max($E_y$)) vs. time plot for perturbation mode $m=1, 2$, domain size $L_x=2,4$ and shear flow parameters $a_v=2a_B$, $v_0 = 1.4v_A$. For $L_x=4$ and $m=1$ case, $E_y(t)$ is calculated as max($E_y(x,z,t)$) at every time inside the blue-colored box shown in Fig. \ref{island}. }
	\label{Ey_plot} 
\end{figure} 
Inspite of strong $m=1$ MHD-KH instability for $L_x=L_z=4$ and $a_v=2a_B$, we found the magnetic island undergo coalescence process by eventually suppressing the unstable MHD-KHI mode generated by super-Alfv\'enic shear flow. This brings out the generality of our findings.}} \\
\section{Summary}
In the present work, using a 2D VR-RMHD model implemented in the BOUT++ framework, we have carried out a systematic study of the effect of in-plane shear flow on the island coalescence problem. Our results in the absence of in-plane shear flow are in very good agreement with previously reported work for our set of parameters. We have applied in-plane shear flows, both parallel and anti-parallel to the magnetic fields. We have calculated the peak reconnection electric field ($E_y$) at the $X$-point for different $v_0$ values keeping $a_v = 2a_B$. To see the effect of the shear length scale, we have calculated $E_y$ for four different values of $a_v$. The main findings are as follows:
\begin{enumerate}
\item When $a_v > a_B$, irrespective of the case whether the system allows the MHD-KHI mode to become unstable or not (decided by $L_z$ value), the shear flow is unable to change the island shape and hence the coalescence process, but significantly reduces the peak $E_y$ or reconnection rate. For our parameter set, $E_y$ decreases by $\sim$70\% as $v_0$ increases from $0$ to $1.4v_A$.

\item For $a_v \leq a_B$, and $v_0 \geq 0.5v_A$ (super-\Alfvenic flow), the MHD-KHI tries to destabilize the magnetic islands shape. But after setting up of plasma circulation, the magnetic islands get stabilized against the strong shear flow and the island coalescence instability dominates over MHD-KHI.

\item {{The plasma circulation inside the islands produces shear flow at the both sides of the reconnection sheet, thus reducing the upstream velocity and hence a reduction in magnetic flux pile-up. With an increase in the value of $v_0$, the plasma circulation becomes stronger.}}


\item Anti-parallel shear flows have the same effect on the current islands as the parallel shear flows (hence not shown).
\end{enumerate} 
\indent
The present study is confined to a single uniform resistivity value for which the plasma is predominantly collisional. Hence the two-fluid effects (Hall physics) and kinetic effects (FLR effects) are safely ignored. In the case of TMI, with slab geometry, several authors have reported a quadrupolar out-of-plane magnetic (say $B_{||}$) field induced by out-of-plane shear flow \cite{Bian2007}. Hall physics also generates quadrupolar $B_{||}$ because of Hall electric field \cite{Bard2018}. Hence, strong out-of-plane flows distort the Hall-induced $B_{||}$ and generate secondary islands \cite{JWang2012}. Also, in the past, for 3D cylindrical geometry, the effect of axial and helical flows on resistive TMI has been shown to be important \cite{Chandra2015}. Hence, we believe it would be very interesting to study the effect of out-of-plane flows and helical flows including Hall physics and kinetic effects on 2D island structure as well as 3D flux tubes. These problems will be addressed in the future.

\section*{Acknowledgment}
The simulations are performed on the Antya cluster at the Institute for Plasma Research (IPR). We would like to thank Dr N. Bisai, IPR for his valuable inputs.
\appendix
\renewcommand\thefigure{\thesection.\arabic{figure}}
\section*{Data Availability}
The data that support the findings of this study are available from the corresponding author upon reasonable request.
%

\end{document}